\documentclass[aps,pre,twocolumn,showpacs,floatfix,showkeys,superscriptaddress,byrevtex,10pt]{revtex4-1}
\usepackage{amsfonts,amssymb,amsmath,bm}
\usepackage{graphicx,epsfig,color}

\DeclareMathOperator\erfc{Erfc}
\begin{document}

\title{A point-particle method to compute diffusion-limited cellular uptake}

\author{A. Sozza} 
\affiliation{Department of Physics, Universit\`a di Torino \& INFN, 
via P. Giuria 1, 10125 Torino, Italy}

\author{F. Piazza}
\affiliation{Centre de Biophysique Mol\'eculaire, CNRS-UPR 4301 and
Universit\'e d'Orl\'eans, F-45071 Orl\'eans Cedex, France}

\author{M. Cencini} 
\affiliation{Istituto dei Sistemi Complessi, CNR,
  via dei Taurini 19 Roma, Italy \& INFN Sezione di "Tor Vergata"
  Roma}

\author{F. De Lillo} 
\affiliation{Department of Physics, Universit\`a di Torino \& INFN, 
via P. Giuria 1, 10125 Torino, Italy}

\author{G. Boffetta} 
\affiliation{Department of Physics, Universit\`a di Torino \& INFN, 
via P. Giuria 1, 10125 Torino, Italy}

\date{\today}

\begin{abstract}
We present an efficient point-particle approach to simulate 
reaction-diffusion processes of spherical absorbing particles 
in the diffusion-limited regime, as simple models of cellular uptake. 
The exact solution for a single absorber is used to calibrate the method, 
linking the numerical parameters to the physical particle radius and 
uptake rate. We study configurations of multiple absorbers of 
increasing complexity to examine the performance of the method, by comparing 
our simulations with available exact analytical or numerical results. 
We demonstrate the potentiality of the method in resolving the complex diffusive 
interactions, here quantified by the Sherwood number, measuring the uptake rate 
in terms of that of isolated absorbers. We implement the method in a pseudo-spectral 
solver that can be generalized to include fluid motion and fluid-particle interactions. 
As a test case of the presence of a flow, we consider the uptake rate by a particle 
in a linear shear flow. Overall, our method represents a powerful 
and flexible computational tool that can be employed to investigate many
complex situations in biology, chemistry and related sciences.
\end{abstract}



\maketitle

\section{Introduction}

Reaction diffusion processes are ubiquitous in many contexts ranging
from physics and chemistry to engineering \cite{rice1985}. They are also
key in biology, where they control enzyme catalysis, antigen-antibody
encounter, fluorescence quenching, and cellular nutrient 
uptake~\cite{berg1993,karp1996,kiorboe2008}, which serves as the main
motivation for this paper. Nutrient uptake typically takes place in a
fluid: flow can therefore modify the reaction rates~\cite{leal2007advanced,neufeld2009chemical}. This is
particularly relevant to unicellular organisms, as the presence of
advection (possibly in combination with motility) modifies the nutrient concentration field and thus the
uptake rate~\cite{karp1996}. In recent years the interest towards the
problem of chemical reactions involving self-propelled bodies in a
flow has increased also due to the technological advancements in
chemically-powered micro/nano-swimmers~\cite{fournier2005synthetic,ozin2005dream,golestanian2007designing}.\\
\indent Here we focus on the widespread diffusion-limited reactions~\cite{rice1985}, 
corresponding to the limit of reactions whose chemical step proceeds
much faster than the diffusive transfer of the
components. Cellular uptake by a spherical cell of radius $R$ can be
approximated~\cite{berg1993} by imposing perfect absorbing conditions
at the particle surface (i.e. vanishing concentration field $\rho$ 
on the sphere's surface). For an isolated
spherical cell of radius $R$ much larger than the nutrient's size, 
the stationary reaction (or uptake) rate is
given by the Smoluchowski formula~\cite{smoluchowski1917} $\kappa_s =
4 \pi D R \rho_\infty$, where $D$ is the diffusion constant of the
nutrient field and $\rho_\infty$ the concentration at infinity. When many
absorbing cells are present, diffusive interactions come into 
play~\cite{traytak1992diffusive,traytak1995diffusive,dorsaz2010,lavrentovich2013,galanti2016}.
This problem of nutrient shielding becomes
even more complex in the presence of a flow that transports the
reactant and/or when the cells move autonomously. 
Other complex situations of biological interest include 
the effects of confining, compartmentalization and active transport of
reactants~\cite{Kalay2012,Grima2006}, such as for 
many biochemical reactions occurring within the cell, and the complex
dynamical organization of the plasma membrane~\cite{Kusumi2012}, where 
dynamic clustering~\cite{Sourjik2004,OConnell2006a}, 
lipid-raft association~\cite{Kusumi2012a,Balint2017} and interactions with 
cytoskeletal elements~\cite{Kusumi2012} of receptors are central in regulating 
how ligand binding triggers biochemical signaling cascades~\cite{Shukla2014}. \\
\indent In all these cases one is interested
in quantifying the relative efficiency of the process in terms of the
ratio between the total uptake rate and the bare diffusive uptake rate of
isolated absorbers/receptors, i.e. the Sherwood number $Sh$. For instance, $Sh<1$
is typically an indication of diffusive interactions (i.e. mutual  screening of 
diffusive ligand flux among receptors, leading to destructive 
interference)~\cite{traytak1992diffusive,lavrentovich2013,galanti2016},
while $Sh>1$ can be obtained when the cell moves relative to the surrounding 
fluid~\cite{karp1996}. Clearly, understanding the adaptations leading to (or
induced by) values of Sh differing from 1 is key to deciphering the
life strategies of many unicellular organisms \cite{kiorboe2008}.\\
\indent Advancements in this fields require experimental, theoretical and
computational work coupling fluid dynamics, ruled by the Navier-Stokes
equation, with the reaction-diffusion rules of reactants. In the case of
natural or artificial micro-swimmers, theory and computations must
correctly describe particles that are advected by the flow, modify it
and react with the transported concentration fields. This is a
formidable challenge, as it requires to resolve the dynamics
on many scales, in particular when the fluid is turbulent.\\
\indent In the absence of a flow, several computational methods have been 
developed, based on finite element method~\cite{eun2013influence}, multipole
expansion techniques~\cite{felderhof1982diffusion,muthukumar1982,bonnecaze1991rate},
first-passage Monte-Carlo techniques~\cite{richards1987diffusion,lee1989random,tsao2001rate}. 
In principle, in this case diffusive interactions 
among many different boundaries can be accounted for exactly via re-expansion 
formulae for a wide array of geometries~\cite{traytak2009,morse1946}.
Recently for example, translation addition theorems for solid spherical
harmonics have been used to compute the reaction rate of diffusion-influenced
reactions~\cite{galanti2016} and investigate transient heat
transfer~\cite{Gordeliy2009} in the presence of many spherical boundaries.
These theoretical treatments have the advantage that in many cases simple
analytical estimates can be obtained by truncating the associated multipole
expansions. For example, when the majority of boundaries are absorbing, simple
monopole approximations have been shown to yield surprisingly accurate
results~\cite{galanti2016,traytak1992diffusive,traytak1995diffusive}.\\
\indent Conversely, for the problem of nutrient
uptake in the presence of a flow there are fewer numerical investigations.
Recent works have studied the uptake of nutrients by active swimmers
in a thin film stirred by their motion~\cite{brandt} and by
diatom chains in two-dimensional flow~\cite{fauci}.
These studies have generalized
the immersed boundaries method (IBM)~\cite{peskin2002immersed} to
account also for the reaction process. IBM converts the no-slip
boundary condition at the body (of the particle or of other
structures) into a set of forces applied on the fluid in the
neighborhood of particle surface so to ensure that the boundary
conditions are fulfilled. In the same spirit the boundary conditions
on the nutrient concentration field are imposed in terms of
appropriate sinks around the particle~\cite{brandt}.  When considering
many (possibly swimming) particles in a stirred fluid, potentially 
turbulent, the above methods become too complex to be used
unless limiting the number of particles, which need several grid
points to be properly resolved. \\
\indent In this work we present a numerical method for computing the
diffusion-limited uptake of nutrients by  small spherical
particles inspired to the Force Coupling Method (FCM),
introduced by Maxey and collaborators~\cite{Maxey1,Maxey2}.
The basic idea of the FCM is to represent each particle 
by a force distributed over a few grid points.
Notwithstanding these limitations, the method
is numerically very effective and compares well with
analytical~\cite{Maxey2} and experimental results~\cite{Maxey3}.
We extend the FCM to the transport of nutrient, by replacing the
absorbing boundary 
conditions with an effective sink of concentration localized at the
particle position (see~\cite{bhalla} for a similar approach).
This method can be easily implemented in the presence of a flow and 
also for self-propelled particles. 
In this work, however, we mainly focus on the diffusive problem 
and compare the results of the FCM with analytical solutions and 
with the exact solutions obtained by a multipole expansion 
method coupled to re-expansion formulae~\cite{galanti2016}.
In this method, the stationary density field is written formally as 
a sum of as many multipole expansions as there are boundaries 
(and local spherical frames of reference). Then, translation addition theorems 
for solid spherical harmonics~\cite{morse1946} are used to express 
the whole density field on each boundary in turn, so that the appropriate 
local boundary conditions can be imposed as many times as there are boundaries.
We also present preliminary results for a single absorber in the
 presence of a linear shear flow,
leaving the study of more complex flows to future investigations.\\
\indent The material is organized as follows. In Section~\ref{sec:method} we
present the method, its implementation and consider the case of an
isolated spherical absorber to explain how the numerical parameter
should be calibrated in order to reproduce the Smoluchowski
result. Then in Section~\ref{sec:multiple} we show the results of the
numerical method in resolving the diffusive interactions between
multiple absorbers in different configurations. In particular, we
consider two absorbers placed at varying distance. Here we can compare
with an exact analytical theory \cite{galanti2016}, allowing us to discuss the
limitations of the method when the particles are too close, or too
far apart. After that, we consider triads and tetrads of particles. Then we use
the method to study random clusters of absorbing particles, either
filling a sphere or a spherical shell, comparing the results both with
exact numerical calculations and approximate analytical
theories. Finally, we show how the reaction rate is modified in the
presence of a linear shear flow, comparing our results with
approximate theories developed in~\cite{frankel1968heat}. 
In the last section we draw the conclusion and describe some possible applications of our method.

\section{The numerical method \label{sec:method}}

We consider a set of $N$ absorbing spherical particles of radius $R_i$
at positions ${\bm X}_i(t)$ ($i=1,2,\ldots,N$).  The scalar field
$\rho(\bm x,t)$ obeys the diffusion equation with absorbing boundary
condition (i.e. $\rho=0$) on the spheres' surface. 
As discussed in the Introduction, we replace the boundary conditions
at the particle surface by a volumetric absorption process of 
first order localized over regularized delta functions $f({\bm x}-{\bm X}_i)$
centered on the particle positions. Hence the concentration field obeys the equation
\begin{equation}
\partial_t \rho  = D \nabla^2 \rho -\rho \sum_i^{N} \beta_i\,f(\bm x-\bm X_i)\,,
\label{eq:modifieddiffusion}
\end{equation}
where $\beta_i$ is the (constant) volumetric absorption rate of particle $i$.
By making $\beta_i$ to depend on the
concentration, the sink term in (\ref{eq:modifieddiffusion}) can mimic
saturable kinetics of Michaelis-Menten type often used for modeling
cellular uptake~\cite{fauci}. 
In this work, however, we are interested in modeling 
perfectly absorbing spheres, for which a number of results are at hand.
Therefore we take the absorption rate constant and we have to 
determine how $\beta_i$ is related to the effective radius 
of the absorbing sphere.
We remark that our method can be implemented also in the
presence of a flow, by adding the advection term in (\ref{eq:modifieddiffusion}),
and also in the case of self-propelled particles, including the 
fluid-particles interactions \cite{Maxey1,Maxey2}.\\
\indent We integrate the diffusion equation (\ref{eq:modifieddiffusion}) 
by a standard pseudo-spectral method in a cubic domain 
of size $L=2\pi$ consisting of $M^3$ grid-points (with $M$ between $64$ and $256$)
with periodic boundary conditions in all the directions.
Time evolution is computed by using a $2^{nd}$ order Runge-Kutta scheme with
exact integration of the linear term.
The use of periodic boundary conditions make the problem equivalent to 
the case of an infinite periodic cubic lattice of sinks, for which
the total concentration decays in time~\cite{traytak1995diffusive}.
In order to reach a stationary state one can add a source term 
to (\ref{eq:modifieddiffusion}), for example by imposing a fixed
concentration over a large bounding sphere in the computational
domain~\cite{lavrentovich2013}, but this cannot be used in the
presence of a flow.
Another possibility is to add a homogeneous source term to 
(\ref{eq:modifieddiffusion}) as done in~\cite{bhalla}.
Because here we are mainly interested in benchmarking the numerical 
method with known results of isolated absorbers in an infinite volume,
we add no source terms to the equations and perform the simulations 
in condition of slowly decaying nutrient.
Nonetheless, by considering a sufficiently large domain with respect to 
the absorber configurations, the effects due to periodicity appear 
only at long times and, as we will see, do not limit the possibility
to measure the nutrient uptake in conditions equivalent to the 
infinite domain.\\
\indent There are several possibilities to implement the regularized delta
function $f(\bm x)$.  For instance, for particle-flow interaction a
Gaussian function is typically employed \cite{Maxey1,Maxey2}.  The
Gaussian, however, has not a compact support and thus is numerically
not very convenient. Here, we adopt a computationally more efficient
choice inspired to immersed boundary methods
\cite{peskin2002immersed}.
We consider the discretized delta function $f(\bm x)$ as the product 
of identical one-variable functions $\phi(x)$
rescaled with the mesh size $\delta x=L/M$ (where $M$ is the number of
grid points):
\begin{equation}
f({\bm x})= \dfrac{1}{\delta x^3} \phi\left(\dfrac{x}{\delta x}\right) 
\phi\left(\dfrac{y}{\delta x}\right) \phi\left(\dfrac{z}{\delta x}\right)\,,
\label{eq:deltareg}
\end{equation}
where $x,y,z$ are Cartesian coordinates. 
The function $\phi$ is chosen symmetric, positive, with a
compact support around its center and normalized. 
The numerical
implementation of (\ref{eq:modifieddiffusion}) requires the evaluation
of (\ref{eq:deltareg}) on a discrete number of points spaced by
$\delta x$. A convenient choice of $\phi$, which is normalized
independently of the number of support points and of the position of
the center relative to the grid (i.e. approximately
grid-translational invariant), is \cite{peskin2002immersed}
\begin{equation}
\phi\left(\frac{x}{\delta x}\right) = \left\{\begin{array}{ll}
\frac{1}{n}\left[1+\cos\left(\frac{2\pi x}{n\delta x}\right)\right] 
& \left|\frac{x}{\delta x}\right| \le \frac{n}{2}\\[0.2cm]
0 & \mbox{otherwise}
\end{array}\right.
\label{eq:halfcosine}
\end{equation}
The particle has a ``numerical radius'' given by $a=(n/2)\delta x$,
which is in general different from its effective radius $R$, i.e. the
radius of the equivalent absorbing sphere, which will (as shown below)
depend on both $a$ and $\beta_i$.  Note that the particle position
${\bm X}_i$ in (\ref{eq:modifieddiffusion}) takes real values in 
three-dimensional space.  Consequently, the smoothed delta function is
centered at any arbitrary position but the function itself is
evaluated only on $n^3$ grid points.\\
\indent The uptake rate $\kappa_i$ of particle $i$ can be directly computed  from 
the integral of the sink term in (\ref{eq:modifieddiffusion})
\begin{equation}
\kappa_i(t) = \int \beta_i\, f({\bm x}-{\bm X}_i) \rho({\bm x},t) d^3{\bm x}
\label{eq:rateint}
\end{equation}
where the integral is numerically evaluated by the sum over the 
grid points defined in (\ref{eq:halfcosine}). The global uptake rate
is then obtained by summing (\ref{eq:rateint}) over all the particles,
or alternatively measuring the rate of change of the volume averaged
concentration $C(t) \equiv \langle \rho \rangle=V^{-1} \int \rho({\bm
  x},t) d^3 {\bm x}$. By integrating (\ref{eq:modifieddiffusion})
is easy to see that
\begin{equation}
\frac{d C}{dt}  = -\frac{1}{V} \sum_{i=1}^N \kappa_i(t)\,.
\label{eq:totalrate}
\end{equation}
%

\subsection{Calibration of the numerical method\label{sec:calibration}}

\begin{figure}[t!]
\centering
\includegraphics[width=0.49\textwidth]{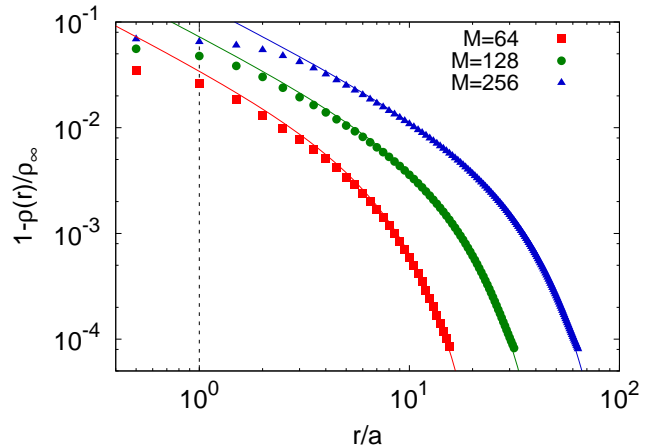}
\caption{
(Color online) Radial profile of the concentration field $\rho(\bm x,t)$ 
obtained from simulations at time $t=5\times 10^{4}\tau_D$ 
at varying the resolution as in legend 
and compared with the theoretical prediction (\ref{eq:app1.2}) (solid lines).
Parameters: $D=10^{-2}$, $\beta=10^{-3}$.}
\label{fig:singleprof}
\end{figure}

\begin{figure}[t!]
\centering
\includegraphics[width=0.49\textwidth]{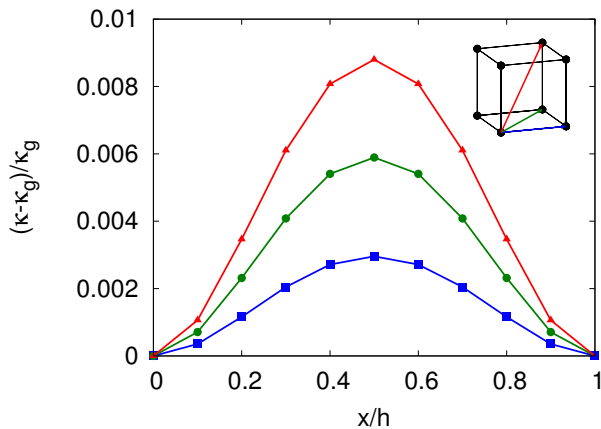}
\caption{(Color online) Relative error on the uptake rates depending 
on the position of the particle within the lattice unit. 
We vary the distance $x$ from a chosen grid point, along three different 
paths (shown in the inset) of size $h$, from bottom to top: 
side of the grid unit (blue, $h=\delta x$), face diagonal (green, $h=\sqrt{2}\delta x$) 
and cube diagonal (red, $ h=\sqrt{3}\delta x$). 
On the $y$-axis we show the discrepancy between the off-grid uptake rate $\kappa$ 
and the one measured on a grid point $\kappa_g$. 
Along each path, the discrepancy is maximal at the farthest position 
from the grid. Simulations refer to resolution $M=64$.}
\label{fig:outgrid}
\end{figure}

In this section we show how the effective radius $R$ of an absorber
depends on $\beta$ and the numerical radius $a$. To this aim we
perform a set of numerical simulations considering a single absorbing
sphere in an initially uniform scalar field, $\rho(\bm x,0)=\rho_0$,
for different values of the absorption rate $\beta$. In all
simulations we fixed $n=4$ in (\ref{eq:deltareg}) (as customary in 
IBM~\cite{peskin2002immersed}), the scalar diffusivity $D=0.01$ and
$\rho_0=1$.\\
\indent The effective radius can be obtained comparing the absorbing rate with
the Smoluchowski result. More precisely, since our simulations are
time-dependent as explained above, one needs to compare the time evolution
of the uptake rate (\ref{eq:rateint}) with the Smoluchovski solution
of the time-dependent diffusive problem (see Appendix~\ref{app1}):
\begin{equation}
\kappa_s(t)=4\pi D R \rho_\infty  \left(1+\frac{R}{\sqrt{\pi D t}}\right)\,.
\label{eq:smolunsteady}
\end{equation}
We use the same symbol for both the time-dependent and the steady 
solution, for the latter $k_s=4\pi D R \rho_\infty$,
obtained from (\ref{eq:smolunsteady}) when $t\to \infty$, the
time dependence is omitted. \\
%
\begin{figure}[t!]
\centering
\includegraphics[width=0.49\textwidth]{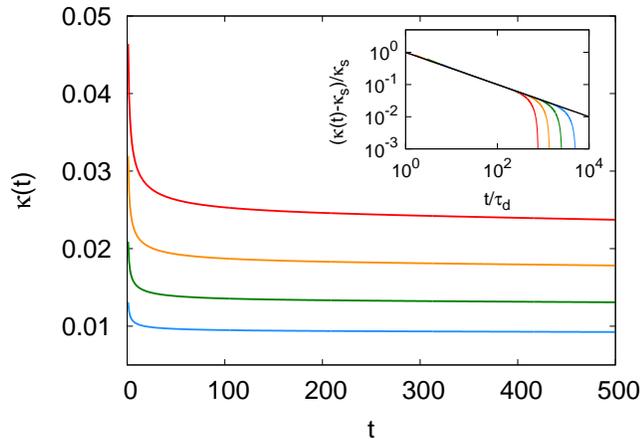}
\caption{(Color online) Time evolution of the uptake rate $\kappa(t)$
for different values of $\beta$, from top to bottom: $2$ 
(red), $2 \times 10^{-1}$ (orange), $5 \times 10^{-2}$ (green) and $2\times10^{-2}$ 
(cyan). Inset: $(\kappa(t)-\kappa_s)/\kappa_s$ versus $\tau=t/\tau_D$ 
($\tau_D=R^2/(\pi D)$ with $R$ obtained from the fit based on eq.~(\ref{eq:smolunsteady})). 
For short times all the curves collapse on the line $1/\sqrt{\tau}$ 
(black line) as predicted by (\ref{eq:smolunsteady}). 
Simulations have been performed with resolution $M=64$.}
\label{fig:kappa}
\end{figure}
%
\indent Figure~\ref{fig:kappa} shows the evolution of the uptake rate
$\kappa(t)$, computed from (\ref{eq:rateint}), as a function of time
for different values of $\beta$.  Two regimes are observed: at the
beginning the diffusive regime described by (\ref{eq:smolunsteady}) is
well evident (see inset), while for longer times a slower decay due to
the boundedness of the domain sets in.  By fitting $\kappa(t)$ with the
expression (\ref{eq:smolunsteady}) in the first regime, one obtains
two independent estimated of $R$ (from the constant term and from the
time dependent term). For all values of $\beta$ the two measures give
the same value of $R$ within $2 \%$ of error.\\
%
\begin{figure}[t!]
\centering
\includegraphics[width=0.49\textwidth]{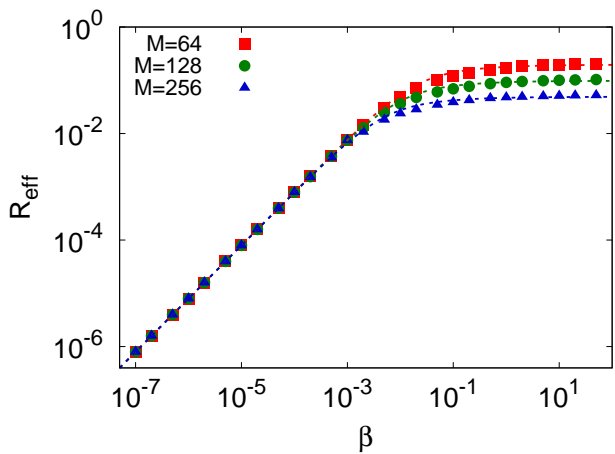}
\caption{(Color online) Effective radius $R$ obtained from $\kappa(t)$ 
fitted with (\ref{eq:smolunsteady}) versus $\beta$ for  different resolutions as labelled. 
The dashed curves display  Eq.~(\ref{eq:calibration}) with $a=(n/2)\delta x$.}
\label{fig:r_beta}
\end{figure}
%
\indent The result of the calibration for the effective radius $R$ is shown in
Fig.~\ref{fig:r_beta} for different resolutions $M$.  For small $\beta$, the
effective radius $R$ is proportional to the absorption rate. For large
$\beta$, $R$ saturates to $a=(n/2)\delta x$ that depends on the
resolution as $n$ is fixed and the mesh size changes as $\delta
x=L/M$.\\
\indent To rationalize this behavior and eventually find an analytical fitting
expression for $R$, we resorted to a crude approximation for the
regularized delta function assuming a spherical sink function of
radius $a$, in polar coordinates $f(r)=\Theta(a-r)/V_a$, where
$\Theta$ is the Heaviside step function, $r$ the distance from the
sphere center and $V_a=4\pi a^3/3$ the sphere volume. With this form
for $f(r)$ the stationary solution of (\ref{eq:modifieddiffusion}) in
the infinite space is easily solved (see Appendix~\ref{app2}).  The
analytic expression of the uptake rate (see Eq.~(\ref{eq:app2_rate})),
when compared with the Smoluchowski rate, yields
\begin{equation}
R = a - \sqrt{\dfrac{D V_a}{\beta}} \tanh \left(a \sqrt{\dfrac{\beta}{D V_a}} \right) \,,
\label{eq:calibration}
\end{equation}
which shows a remarkable (within $5\%$) agreement with
 the effective radius obtained with the fitting
procedure (see Fig.~\ref{fig:r_beta}). Thus Eq.~(\ref{eq:calibration}) can
be used as the calibrating function. Notice that for small $\beta$
Eq.~(\ref{eq:calibration}) yields $R= \beta/(4\pi D)$, which is the
result one would obtain by replacing $f$ with a true $\delta$-function in
(\ref{eq:rateint}), while $R\to a$ for large $\beta$.
Typically, in our simulations we fixed $\beta=D=0.01$ 
which leads to an effective radius $R\lesssim a$.\\
\indent It is worth remarking that, as in the case of the Force Coupling
Method for fluid-particle interaction~\cite{Maxey1,Maxey2}, the
diffusive boundary layer is not well resolved at the scale of the
regularization. This is apparent in Fig.~\ref{fig:singleprof}, which
displays the profile of the scalar field as a function of the
distance from the particle. The analytical expression obtained from
Eq.~(\ref{eq:app1.2}) agrees well with the numerical result only for
$r > 2 a$, which depends on the resolution, even if $R\ll a$.  This,
as we will see in the next section, has some repercussions on the
ability of the method to resolve the diffusive interactions of close
particles.\\
%
%
\indent Finally, in Figure \ref{fig:outgrid}, we assess possible systematic
errors coming from varying the position of the particle in the grid,
i.e. errors due to the use of the regularized $\delta$ function
(\ref{eq:deltareg}). We measured the relative error on the uptake rate
varying the position of the particle along the side $\delta x$ of a
lattice unit, along the face diagonal and along the cube diagonal (see
inset). The largest variation observed was less than $1\%$ for a
regularization on $n^3=64$ grid points.\\
%
%

\section{Configurations with multiple absorbers\label{sec:multiple}}

In this Section we consider $N$ static absorbing particles, arranged
in configurations of increasing complexity from regular to random,
with the aim of testing the reliability and precision of our
method. For the sake of simplicity, we discuss only cases in which all
particles have the same radius , i.e. $\beta_i=\beta$ in
Eq.~(\ref{eq:modifieddiffusion}). 
All simulations are initialized with a
uniform scalar field, $\rho(\bm x,0)=\rho_0=1$. The asymptotic uptake
rates are evaluated as discussed in Sect.~\ref{sec:calibration} by
fitting $\kappa(t)$ with (\ref{eq:smolunsteady}) on each particle.
Indeed, it can be shown that the functional form (\ref{eq:smolunsteady})
holds also in the case of multiple sinks 
\cite{traytak1992diffusive, traytak1995diffusive}.\\
\indent We compare the numerically obtained rates with 
those obtained from a numerical multipole expansion algorithm~\cite{galanti2016}.
When available, we also compare our results with analytical exact or
approximate expressions.  The main aim of this study is the
validation of our method in resolving the diffusive interaction,
quantified by the Sherwood number $Sh$ defined as the total absorption
rate normalized with that of $N$ isolated absorbers
\begin{equation}
Sh = \dfrac{\kappa_{tot}}{N \kappa_s}\,.
\label{eq:sh}
\end{equation}
In the last subsection we shall consider the case of a single absorber
in the presence of a linear shear flow and study the Sherwood number
as a function of the Peclet number, quantifying the ratio between
advective over diffusive transport.

\subsection{Pairs of absorbers ($N=2$)\label{sec:pairs}}

\begin{figure}[t!]
\centering
\includegraphics[width=\columnwidth]{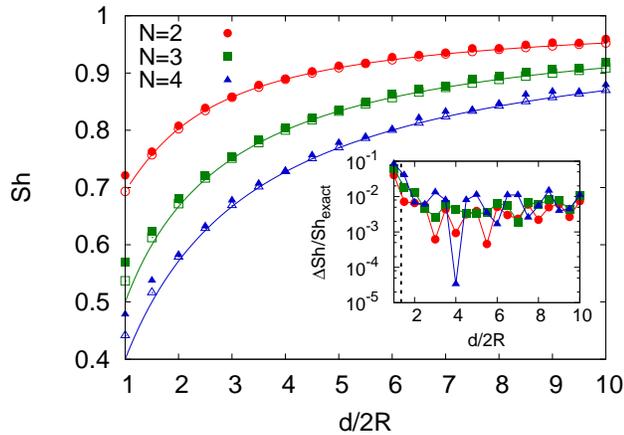}
\caption{(Color online) $Sh$ for configurations of $N$ equidistant
  absorbers as a function of the dimensionless distance $x=d/2R$ for
  $N=2,3,4$ as in legend.  Filled symbols are the results from the
  numerical simulations, open symbols exact results obtained 
  with the method of Ref.~\cite{galanti2016}.
  Lines represent the monopole
  approximation eq.~(\ref{eq:sh_reg}).  The inset shows the relative
  error with respect to the exact solutions.  The vertical dashed line is
  positioned at $x_m = d_m/2R$, marking the condition
  for well-resolved simulations.  On the left of the threshold line,
  the error rises to $4-8\%$ while for the other distances it is
  always close to $1\%$.  The resolution used is $M=128$.}
\label{fig:regular}
\end{figure}

The case of two spherical absorbers of radius $R$ separated by a
distance $d$ is one of the few examples of diffusive interaction
problem that can be  solved exactly.  After choosing bi-spherical
coordinates, the Laplace equation becomes separable~\cite{morse1946}
and the total absorption rate depends on the relative distance
$x=d/2R$ as~\cite{samson1999,piazza2005}
\begin{equation}
Sh = \sqrt{x^2-1}\sum_{n=0}^\infty\frac{2}{1+(x+\sqrt{x^2-1})^{2n+1}}\,.
\label{eq:pairsseries}
\end{equation}
In the limit of well-separated absorbers, $x \to \infty$,
Eq.~(\ref{eq:pairsseries}) yields the non-interacting result $Sh=1$
(i.e. both spheres absorb the nutrient at the Smoluchovski rate as if
they were isolated). Notice that, already for $x\gtrsim 2$ the first
correction given by the monopole contribution,
\begin{equation}
Sh=\dfrac{2 x}{2 x + 1} = \dfrac{d}{d + R}\,,
\label{eq:pairsmoa}
\end{equation}
is a very good approximation of (\ref{eq:pairsseries}).  In
the limit of two spheres in contact, $x=1$, Eq.~(\ref{eq:pairsseries})
gives the maximum interference, with $Sh=\ln(2)$.\\
%
\begin{figure}[t!]
\centering
\includegraphics[width=0.49\textwidth]{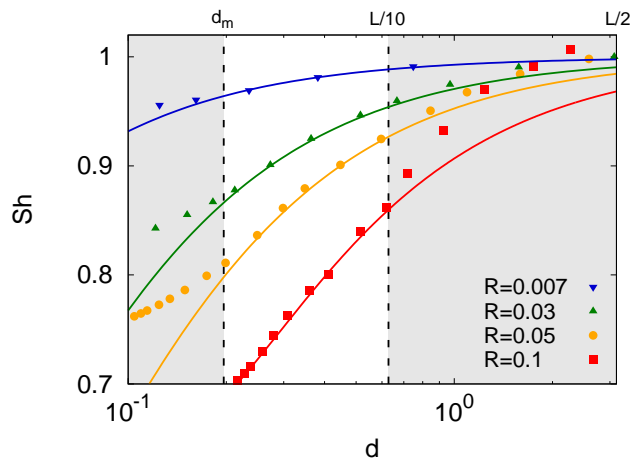}
\caption{(Color online) $Sh$ as a function of the distance $d$ between 
two spherical absorbers of radius $R$, varied as in legend. 
Solid curves represent the exact result (\ref{eq:pairsseries}). 
The grey shaded regions highlight the range of distances where the 
numerically obtained $Sh$ becomes sensibly different from the exact 
value, see text for discussion. 
The simulations have been performed with $M^3=64^3$ grid points.}
\label{fig:pairs}
\end{figure}
%
\indent In Fig.~\ref{fig:pairs} we show the numerically computed Sherwood
number as a function of the pair distance $d$ for different choices of
the particle effective radius $R$ at fixed resolution. The numerical results
are directly compared with the exact value obtained from
(\ref{eq:pairsseries}). A very good agreement between numerical and
theoretical values is observed for distances larger than $d_m \approx
a=(n/2)\delta x$,  and smaller than, about  $1/10 \div 1/8$ of the domain
size $L=2\pi$.\\
\indent The discrepancies at small distances are due to the fact that the
method does not resolve the particle surface: the numerical radius,
$a$, imposed by the regularized delta function turns out to be the
limiting distance for resolving the pair diffusive interactions (see
also Fig.~\ref{fig:singleprof} and related discussion), regardless of
the effective radius of the particles. To reduce $d_m$ the only
possibility is thus to increase the resolution.\\
\indent The large-distance discrepancies are due to the periodicity of the
simulation domain. Since the diffusive interactions are long-ranged
(they decay with the inverse of the distance from the absorbers, see
Eq.~(\ref{eq:app1.2})), when $d$ increases the particles start to
interact not only with each other but also with their periodic
images, leading to an increase of the total uptake rate 
(i.e. $Sh$ becomes larger than predicted for a pair of absorbers 
at the same distance in the infinite space). 
This effect tends to increase with the effective radius $R$ of
the particle as the diffusive interaction increases with $R$ (see
Eq.~(\ref{eq:app1.2})). We emphasize that this effect cannot be
modified by changing the resolution but requires working with
different boundary conditions.\\
\indent Summarizing, the above results show that provided the particles are at
distances $(n/2)\delta x = d_m < d < d_M \approx L/10-L/8 $ the
numerical method works quite well.  Figure~\ref{fig:regular} shows
$Sh$ as function of the rescaled distance $x=d/2R$ together with the
exact result and the monopole approximation. As it is clear from the inset,
in the whole range of $x$ the numerical values are within $2\%$
from the exact result with larger deviations, $\sim 4\%$, when $x\to
1$ corresponding to distances $d \approx d_m$. In the following we
shall exploit these results when studying more complex arrangements of
absorbing particles, making sure that the particles stay at distances
within the range of scales for which the method works well.
%
%

\subsection{Regular triangles ($N=3$) and tetrads ($N=4$)}

We now consider regular arrangements of $N=3$ and $4$ particles at
varying  distances. From the theoretical side, the monopole
expression (\ref{eq:pairsmoa}) can be easily extended to the case of $N>2$
absorbers.  Within the monopole approximation, one
can write the set of linear equations for the uptake rate $\kappa_i$
of the $i$-th absorber
\cite{traytak1992diffusive,traytak1995diffusive}
\begin{equation}
\kappa_i = \kappa^i_s - \displaystyle\sum_{j\ne i}^N \epsilon_{ij} \kappa_j, 
\qquad i=1,...,N,
\label{eq:moa}
\end{equation}
where $\epsilon_{ij} = R_i/d_{ij}$, $R_i$ is the radius of the
$i^{th}$ sphere, $d_{ij}$ the distance between the $i$-th and $j$-th
sink and $\kappa^i_s=4 \pi D R_i \rho_\infty$.  The case considered
here is $R_i=R$ and $d_{ij}=d$, for $N=3,4$. In this limit, the total
Sherwood number, in the monopole approximation, is given by
($x=d/2R$):
\begin{equation}
Sh = \dfrac{2 x}{2x+ (N-1)}\,.
\label{eq:sh_reg}
\end{equation}
The numerical results are shown in Figure~\ref{fig:regular}, together
with the monopole expression~(\ref{eq:sh_reg}) 
and the exact results
computed by using the approach described in Ref.~\cite{galanti2016}. 
The limit $x=1$ corresponds to the minimal distance at
which the spheres are at contact and therefore to the maximum
diffusive interaction.  From a numerical standpoint, with our choice of
$\beta$, this limit corresponds to $d\approx d_m$. The discrepancy
between the numerical simulations and the exact results is here
maximal, between $4\%$ and $8\%$ (see inset) increasing with $N$ as
intuitively expected. At larger distances  the exact values are
recovered within $\lesssim 2\%$. It is remarkable that the interaction
is still observed for $x \simeq 10$, as a consequence of the long-range 
nature of the diffusive interactions. Notice that, to reach
$x=10$ without violating the constraint imposed by the periodic
boundary conditions (cfr. Fig.~\ref{fig:pairs}), we have varied $x$
also changing the effective radius. We finally remark that as soon as
$x>2$ the monopole approximation practically coincides with the exact
result.

\subsection{Deformed Triangles $N=3$}

%
\begin{figure}[t!]
\centering
\begin{tabular}{cc}
\includegraphics[width=0.49\textwidth]{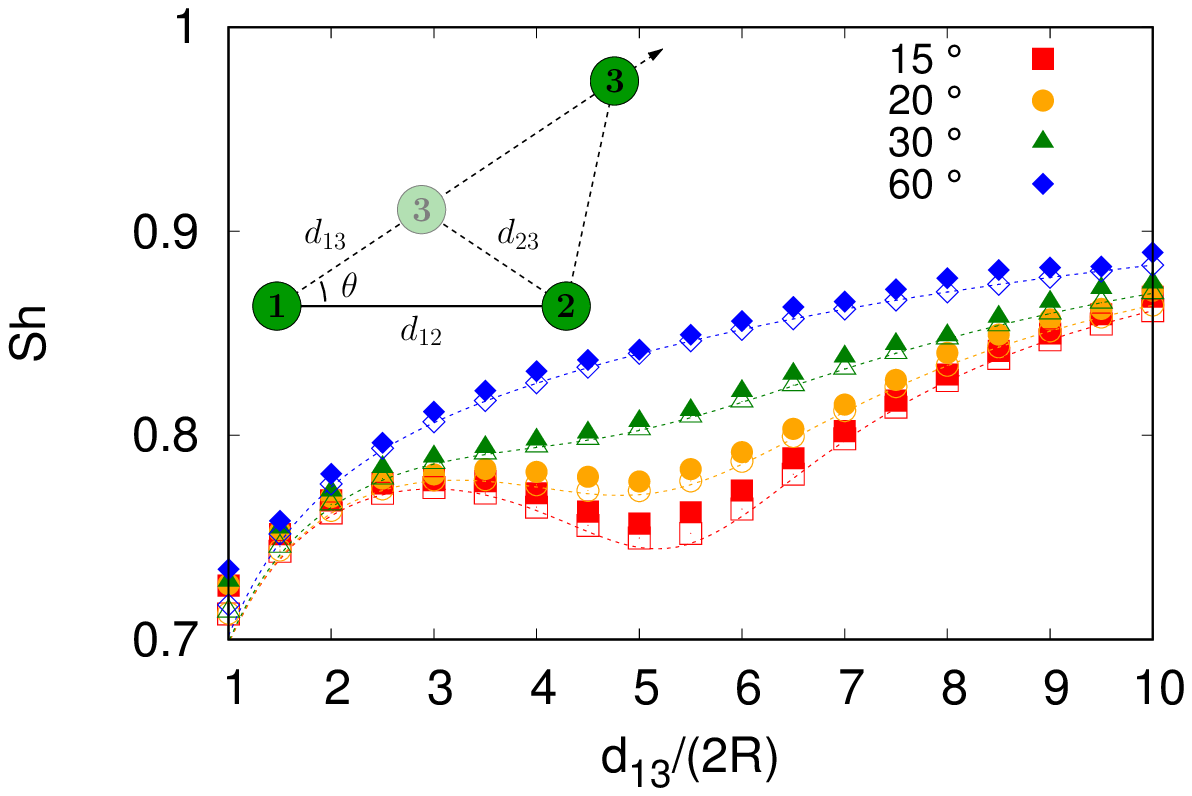}\\
\includegraphics[width=0.49\textwidth]{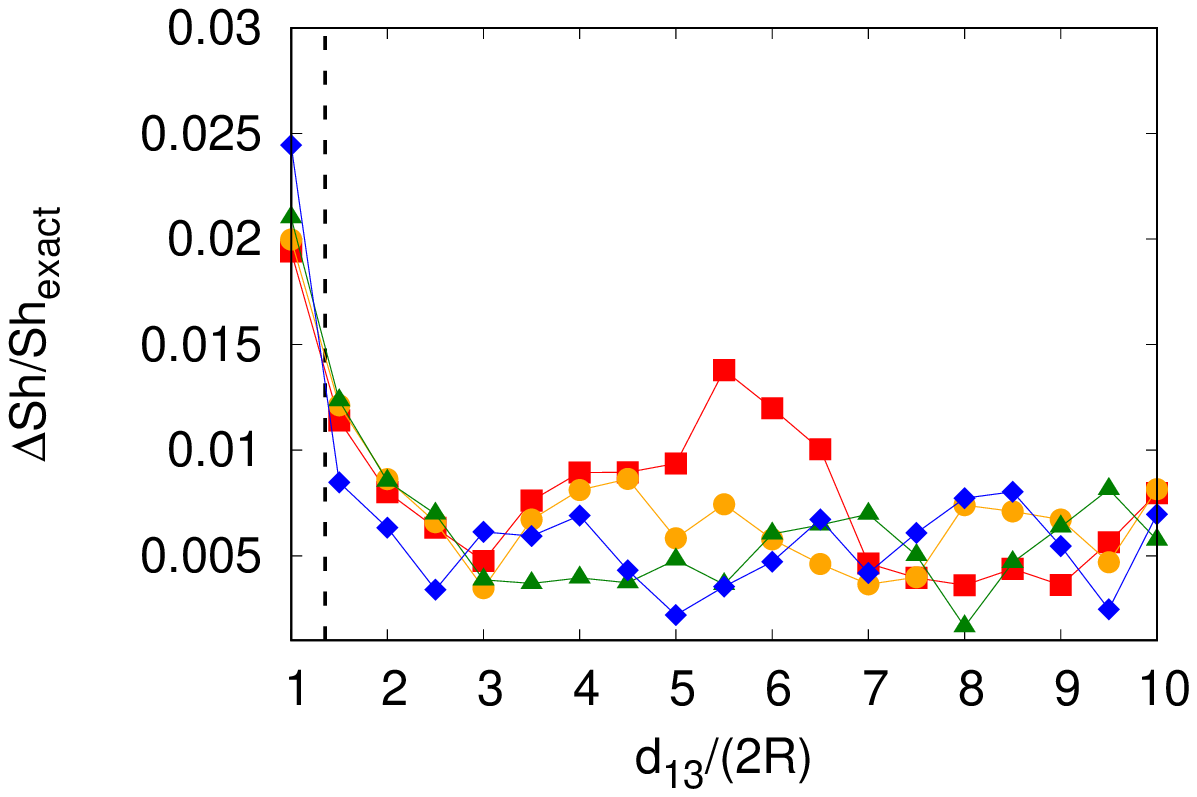}
\end{tabular}
\caption{(Color online) Top Panel: $Sh$ for different configurations 
of deformed triangles with $\theta=15^\circ,20^\circ,30^\circ,60^\circ$ 
as a function of the distance $d_{13}$ normalized by the sphere diameter $2R$. 
The minimum of $Sh$ corresponds to the configuration of minimal distance $d_{23}$ 
(filled symbols from the numerical simulations, 
open symbols obtained with the method of Ref.~\cite{galanti2016}). 
Inset: Sketch of the deformed triangles. 
Bottom Panel: relative error of the simulations with respect to the exact results 
obtained with the method of Ref.~\cite{galanti2016}. 
The dashed line represents the critical distance $x_m = d_m /(2R)$ 
below which the numerical accuracy is decreased as discussed in Sect.~\ref{sec:pairs}.}
\label{fig:triads}
\end{figure}
%
We considered also the case of three spheres of
radius $R$ at the vertices of irregular triangles as sketched in the
inset of the top panel in Fig.~\ref{fig:triads}. A practical way to
construct the triangle is the following: We fix the distance between
particle 1 and 2 to be $d_{12}=X d_m$ with $X>1$. Let us denote with $\theta$
the angle between the segments $12$ and $13$.  We keep  this angle fixed and
vary the distance $d_{13}=d$, requiring $d_{ij}\geq d_m$ for $i\neq j =1,2,3$,
which implies a minimal angle $\theta_c=\arcsin(1/X)$.  Here we fix $X=5.5$ so
that $\theta_c \sim 14.35$. In the simulations we used
$\theta=15^{\circ},20^{\circ},30^{\circ},60^{\circ}$ and varied $d$ in
the range $\in[2R,20R]$.  As for the parameters of the simulation we
fix $D=\beta=0.01$ in such a way that the radius of the spheres $R=0.036$.\\
\indent The solution of such configurations in the monopole approximation
using Eq.~(\ref{eq:moa}) is given by
\begin{equation}
Sh = \dfrac{1}{3}+\dfrac{2}{3}\dfrac{(A-1)(B-1)(C-1)}{A^2+B^2+C^2-2ABC-1},
\label{eq:sh_triads}
\end{equation}
where $A=R/d_{12}$ with $d_{12}$ fixed, $B=R/d_{13}$ and $C=R/d_{23}$ 
with $d_{23} = \sqrt{d_{12}^2+d_{13}^2-2d_{12}d_{23}\cos\theta}$. \\
\indent In Fig.~\ref{fig:triads} we plot the total Sherwood
number of the triadic system  as a function of the distance
$d_{13}$ normalized by the diameter of the absorber $2R$.  Our
simulations, are compared with
the exact results obtained following 
the method of Ref.~\cite{galanti2016}.
 We also compare the results with the monopole approximation
(\ref{eq:sh_triads}).  The minimal uptake is obtained in the
configuration with minimum distance $d_{23}$, which maximizes the
diffusive interactions.  As shown in bottom panel the error is within
$2\%$, but for configurations with $d_{13}\approx d_m$, as expected
from previous discussions. We conclude by noticing that when particle 3
is moved far away from the pair, we recover the asymptote (not shown)
given by the uptake of a single sphere and the contribution of the
pair alone, which with our choice is $Sh\simeq 0.94$.
%

\subsection{Random Spherical Cluster}

\begin{figure}[t!]
\centering
\includegraphics[width=0.49\textwidth]{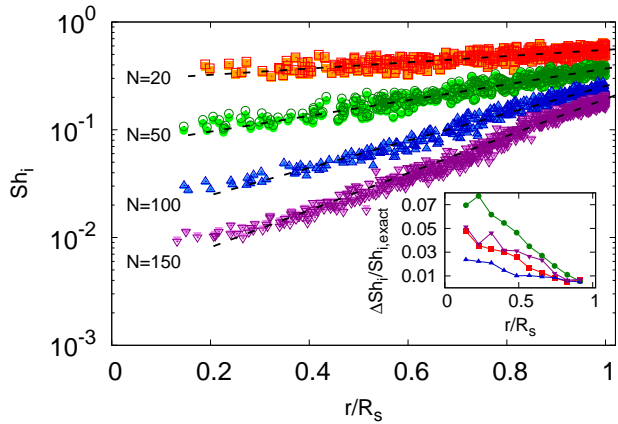}
\caption{(Color online) 
Individual Sherwood number $Sh_i=\kappa_i/\kappa_s$ for absorbers in
a spherical cluster with $N$ particles as labeled. Filled symbols denote the 
numerical results obtained by our numerical method, empty symbols
those obtained with the exact numerical computation described in 
Ref.~\cite{galanti2016}.
Dashed lines represents the prediction (\ref{eq:phi2}).
Inset: relative errors of the uptake computed as the average 
absolute distance of the two numerical methods, plotted as a function 
of the radial coordinate.}
\label{fig:radialsh}
\end{figure}

In this Section we consider the case of a cluster of absorbers. 
One important motivation comes from biology in the case of 
colonies of microorganisms.
In this case one is interested in understanding how
diffusive interactions, which cause nutrient shielding for
cells in the cluster interior, deplete the growth of the colony 
\cite{lavrentovich2013}.\\
%
\begin{figure}[t!]
\centering
\includegraphics[width=0.49\textwidth]{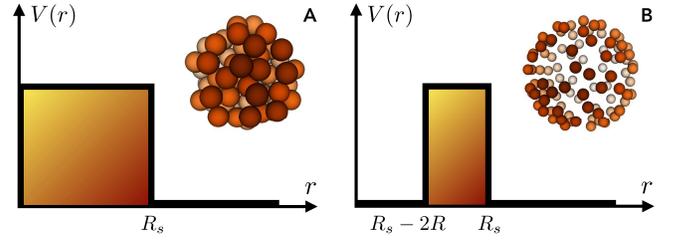}
\caption{Sketch of the cluster configurations: (A) spherical cluster; (B) sherical shell cluster. 
The graphs also show the potential used to develop the effective medium approximation.}
\label{fig:sketch}
\end{figure}
%
\indent Specifically, we consider a spherical cluster of absorbers, i.e. a
sphere of radius $R_s$, centered at the origin, comprising $N$ spherical
absorbers, with the same radius $R$, randomly arranged in its interior 
avoiding geometrical overlaps (see Fig.~\ref{fig:sketch}A). 
In this case it is possible to have an analytical estimation
of the nutrient uptake by introducing an effective-medium approximation
\cite{lavrentovich2013}.\\
\indent The basic idea of the method is to introduce an effective concentration
field $\psi(\bm x,t)=\langle\rho(\bm x,t)\rangle$ where the the brackets denote 
an ensemble average over the possible random position of the absorbing particles. 
By averaging both sides Eq.~(\ref{eq:modifieddiffusion}) 
and assuming stationarity, one has
\begin{eqnarray}\label{eqe}
D \nabla^2 \psi = \left\langle \sum_{i=1}^{N} \beta_i \rho(\bm x)
\delta(\bm x - \bm X_i) \!\!\right\rangle \\[0.2cm]
 \simeq \int \chi(\bm x-\bm x')\psi(\bm x')d\bm x', \nonumber
\end{eqnarray}
where $\chi(\bm x)$ is a linear-response function describing the
deformation of the concentration field induced by the absorption~\cite{lavrentovich2013}.  
The linear-response approximation is only
valid for sufficiently small concentration field deformations (dilute
clusters) and away from the cluster edges.  Fourier transforming
equation (\ref{eqe}), one obtains the equation $-Dq^2 \hat{\psi}(\bm
q) = \hat{\chi}(\bm q)\hat{\psi}(\bm q)$. Since, on average, the cells
are isotropically distribuited, $\chi(\bm q)$ can only depend on
$q=|\bm q|$.  Expanding $\hat{\chi}$ around $q=0$ and truncating at
the zeroth order, i.e.  $\hat{\chi}(0)=k$, provides the desired
mean-field approximation. Hence the configurationally averaged nutrient
concentration obeys the equation $D\nabla^2 \psi - k\psi = 0$
valid within the sphere of radius $R_s$ delimiting the cluster,
outside $D \nabla^2 \psi=0$, this is nothing but the equation we
already solved to determine the calibrating function
(\ref{eq:calibration}) (see Appendix~\ref{app2}). In the above
expression $k$ represents an effective absorption rate within the
sphere in the macroscopic description. The truncation at zeroth order
works reasonably well for dilute clusters, and in this limit
$k=4\pi D R N/V_s = 3 D \phi/R^{2}$ where
$\phi=NR^3/R_s^3=N/\alpha^3$ is the volume fraction (with $\alpha=R_s/R$). 
The cluster is thus approximated as a
unique sink with penetrable walls.
We can directly use
Eq.~(\ref{eq:app2_rate}) to express the total uptake rate
\begin{equation}
  \label{k_eff}
\kappa_{tot} =4 \pi D \psi_\infty \left[R_s - \xi \tanh\left(\dfrac{R_s}{\xi}\right)\right],
\end{equation}
 $\xi=\sqrt{D/k}=R/\sqrt{3\phi}$ being the 
{\it penetration length}.\\
\indent The Sherwood number is defined as $Sh=\kappa_{tot}/N\kappa_s$, 
so using $\kappa_s=4\pi D R \psi_\infty$ and replacing $N=\alpha^3\phi$, 
from (\ref{k_eff}) we obtain
\begin{equation}
Sh(\phi)=\dfrac{1}{\alpha^2\phi}\left[1-\dfrac{1}{\alpha\sqrt{3\phi}}\tanh\left(\alpha\sqrt{3\phi}\right)\right].
\label{eq:totshspherical}
\end{equation}
Let us now consider the local uptake rate of a cell within the cluster. We denote with $\kappa_i$ the
uptake rate of the $i$-th particle and we identify its position in the cluster by its distance $r_i$ form the center. We can then introduce
the average uptake rate $\kappa(r)=\langle \kappa_i \big|r_i=r\rangle$, 
where the brackets indicate the ensemble average over different
configurations. The Sherwood number of a cell at a distance $r$ from the center of the cluster will then be $Sh(r) = \kappa(r)/\kappa_s $. By definition, the total uptake rate of the cluster
is given by $\kappa_{tot} = \langle \sum_i^N \kappa_i \rangle$, while the total flux absorbed by the particles contained in a smaller sphere of radius $r$ is given by
\begin{equation}
4\pi D r^2 \dfrac{d\psi}{d r}
= N\int_0^r \kappa(r') p(r')dr',
\label{eq:fluxr}
\end{equation}
where $p(r)dr=V_s^{-1} 4\pi r^2 \theta(R_s-r) dr$ is the probability to find a particle at a certain radial position.
By taking the derivative of expression (\ref{eq:fluxr}), one gets
\begin{equation}
Sh(r) = 
\dfrac{\xi^2}{\psi_\infty r^2} \dfrac{d}{dr} \left( r^2  \dfrac{d}{dr} \psi \right).
\label{eq:phi2}
\end{equation}
By noting that $\xi^2 \nabla^2 \psi = \psi$, it's easy to see that
$Sh(r) = \psi(r)/\psi_\infty$, i.e. the local uptake rate $\kappa(r)$ is proportional to the averaged concentration profile $\psi(r)$.\\
\indent We turn now to the numerical results. We considered random
distribution of particles in a spherical cluster of radius
$R_s=L/8=\pi/4$ so to minimize effects due to the periodic boundary
conditions (cfr. Fig.~\ref{fig:pairs}). As for the absorbers, we
considered $N=20,50,100,150$ spheres with effective radius
$R\sim0.036$ so that the (nominal) volume fraction ranges in
$\phi=2\cdot 10^{-3}-10^{-2}$, that is, small enough for the effective
medium approximation to be accurate.  Particles are placed uniformly
within the sphere volume, ensuring that they stay at distances larger than
$d_m \approx (n/2)\delta{x}=(n/M)\pi $ to reduce the errors due to poor
resolution of the diffusive interaction at short distances
(Cfr. Fig.~\ref{fig:pairs}). For each $N$ we considered from $5$ to
$10$ different configurations to perform ensemble averages and thus to
compare with the effective field approximation. The same
configurations have been used to 
compute the exact solution with the method of Ref.~\cite{galanti2016}.\\
%
\begin{figure}[t!]
\centering
\includegraphics[width=0.49\textwidth]{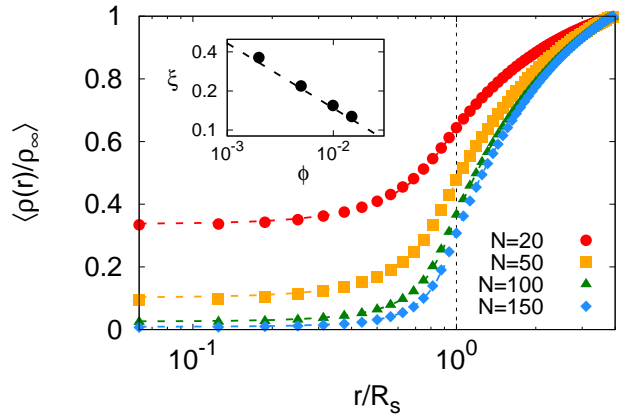}
\caption{(Color online) Average radial concentration profile $\langle\rho(r)\rangle$ vs $r/R_s$, 
for clusters with $N=20,50,100,150$ particles as labeled. 
Numerical data (symbols) are compared with the effective medium solution given by 
eq.~(\ref{eq:app2_solution}) (dashed curves). 
Inset: Penetration length $\xi$ obtained from the fit of the radial profile (filled circles) 
and theoretical prediction given by $\xi=R/\sqrt{3\phi}$~\cite{traytak2013ligand} (dashed line). 
The average is performed over $5-20$ independent realizations. 
The numerical resolution is $M=128$.}
\label{fig:averageprofile}
\end{figure}
%
\indent Figure~\ref{fig:averageprofile} shows the average density profile
$\psi(r)=\langle \rho(r) \rangle$ compared with the theoretical
prediction given by the effective medium approximation 
(\ref{eq:app2_solution}). The agreement is remarkably good.\\
\indent From the simulations, we can extract the uptake rate of each particle
using the standard fitting procedure, which involves the temporal
evolution of uptake rate at intermediate times (see
Sect.~\ref{sec:calibration}).  By averaging over particles and
different configurations, one obtains a measure of the total uptake
rate of the cluster.  Alternatively, one can extract the uptake rate
of the cluster directly from the concentration field $\rho$.  The
radial profile can be compared with the theoretical prediction
(\ref{eq:app2_solution}) to extract the parameters of interest.  From
the inner solution it is possible to extrapolate the penetration length
$\xi$ shown in the inset of Fig.~\ref{fig:averageprofile} with the
scaling $R/\sqrt{3\phi}$.  By fitting the behavior in the outer
region, we have an alternative estimate of the total Sherwood number.\\
\indent In Figure~\ref{fig:radialsh} we plot the individual Sherwood number
$Sh_i=\kappa_i/\kappa_s$ as a function of the radial distance of 
particles in the cluster, compared with the values obtained 
from the exact numerical solution. The relative difference is 
below $8\%$, as shown in the inset, and is larger in the interior
of the cluster, due to the accumulation of errors on the
concentration density due to the outer absorbers.
In Fig.\ref{fig:radialsh} we also plot the theoretical prediction
(\ref{eq:phi2}).\\
\indent Finally, in Fig.~\ref{fig:totsh} we show that also the total Sherwood
number compare very well with the theoretical prediction
(\ref{eq:totshspherical}) and the exact computation (with error within
$10\%$).
%
%

\subsection{Spherical Shell Cluster}

\begin{figure}[t!]
\centering
\includegraphics[width=0.49\textwidth]{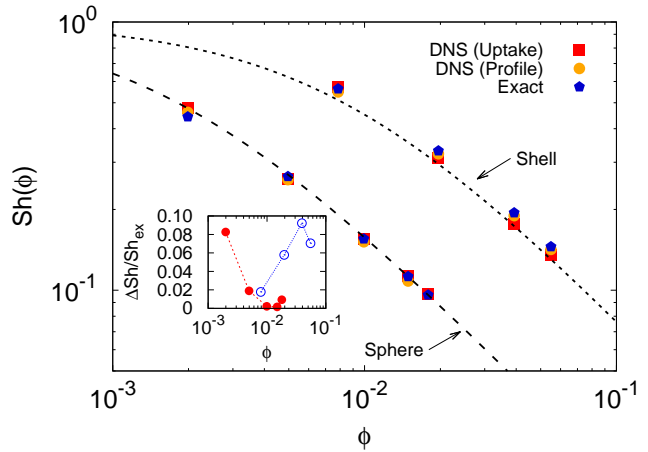}
\caption{Total normalized Sherwood number $Sh=\kappa/N\kappa_s$ for the spherical
  cluster and spherical shell as a function of the volume fraction
  $\phi$, averaged over 5 to 10 configurations and obtained by fitting
  the instantaneous uptake (red squares) or fitting the outer solution
  of the radial concentration profile (orange circles).  Simulations are
  compared with exact results calculated using the method 
  described in Ref.~\cite{galanti2016}
  (blue pentagons). Inset: Relative error
  between simulations and exact results for the spherical cluster (filled circles) 
  and for the spherical shell (empty circles).
\label{fig:totsh}}
\end{figure}

In this section we study a generalization of the spherical cluster,
considering absorbers  with their centers at a
fixed distance from the origin of the sphere of radius $R_s$ (see
Fig.~\ref{fig:sketch}b). This kind of configuration is encountered in
nature. For example, \textit{Volvox} is a colonial alga forming spherical
colonies with a $1$-mm diameter, is usually $100$
times larger than the single cell forming it and contains up to
$5\,10^4$ cells organized as a monolayer of flagellated cells on the sphere surface \cite{miller2010volvox}.\\
\indent Following the same idea used for developing the effective medium
approximation of spherical cluster, one can develop an analytical
description for spherical shell clusters. 
In particular, after averaging over the absorbers configurations 
and performing the expansion of the response function
we end up with the equation:
\begin{equation}
  D\nabla^2{\psi}=k [\Theta(R_s - |\bm r|)+\Theta(|\bm r|-(R_s-2R))] \psi,
  \label{eq:spsh}
\end{equation}
$\Theta$ being the Heavyside step function. The above equation must be
solved with the boundary condition $\psi(r\to \infty)=\psi_\infty$ and
where again $k=3D\phi/R^2$ in the dilute limit. 
Now the volume fraction is given by $\phi=NV_p/V_s$ 
with $V_p=4\pi R^3/3$ the volume of the single absorber 
and $V_s$ the volume of the shell between $R_s-2R$ and $R_s$.
The solution of Eq.~(\ref{eq:spsh}) is detailed in Appendix~\ref{app3}.\\
\indent Similarly to the previous section, introducing $\alpha=R_s/R$ and 
$N=\phi V_s/V_p=2\phi(3\alpha^2-6\alpha+4)$, and using the expression for the total 
uptake rate Eq.~(\ref{eq:app3_rate}), after some algebra, the total
Sherwood number can be expressed as
\begin{small}
\begin{equation}
Sh(\phi) = \dfrac{\frac{1}{2}[\alpha(\alpha-2)\sqrt{3\phi}-1/\sqrt{3\phi}]\tanh(2\sqrt{3\phi})\!+\!1}
{\phi(3\alpha^2-6\alpha\!+\!4)[(\alpha-2)\sqrt{3\phi}\tanh(2\sqrt{3\phi})\!+\!1]}.
\label{eq:totsh_shell}
\end{equation}
\end{small}
In Fig.~\ref{fig:totsh} the total Sherwood number is compared with the
theoretical prediction (\ref{eq:totsh_shell}), the agreement is good
within $10\%$.  It is also interesting to note that the present
configuration in spherical shell-like geometry enhances the uptake
rates per cell and the total uptake rates of the cluster with respect to
the configurations of bulk spherical clusters. Therefore, it can
represent an efficient strategy to maximize the uptake rate.
%
%

\section{Spherical absorber in a linear shear flow}\label{sec:shear}

We end testing the method in the presence of a flow, we
consider a single (non-rotating) absorber of radius $R$ placed in the
position of zero velocity (so that it does not move) of a linear
shear, ${\bm u}=\gamma(y,0,0)$, for which analytical results are
available~\cite{leal2007advanced, karp1996}.  Nutrient evolves
according to Eq.~(\ref{eq:modifieddiffusion}) with the addition of the
advection term:
\begin{equation}
\partial_t \rho  +\bm{u}\cdot\bm{\nabla}\rho = 
D \nabla^2 \rho -\rho \sum_i^{N} \beta_i\,f(\bm x-\bm X_i)\,.
\label{eq:advectiondiffusion}
\end{equation}
Analytical results predicts that for small Peclet numbers
\footnote{N.A. Frankel and A. Acrivos (1968) attain the result 
$Nu=2+c' \,Pe'^{1/2}$, by defining $Nu$ (the homolog of $Sh$ for the heat transfer) 
based on the particle diameter, $Pe'=2 R^2\gamma/D$ and with $c'=0.9104/\sqrt{2\pi}\simeq 0.36$.
Transforming in our variables $Nu=2\,Sh$ and $Pe' = 2\,Pe$, we obtain 
the relation $Sh=1+c\,Pe^{1/2}$, with $c = c'/\sqrt{2} \simeq 0.26$.}, 
$Pe=\gamma R^2/D$ ,
the Sherwood number behaves as~\cite{frankel1968heat,karp1996}
\begin{equation}
Sh \approx 1 + 0.26 Pe^{1/2}\,.
\label{eq:match}
\end{equation}
%
\begin{figure}[t!]
\centering
\includegraphics[width=0.49\textwidth]{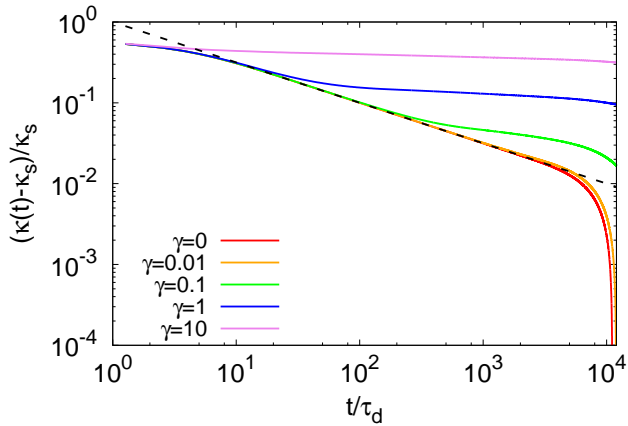}
\caption{(Color Online) Time evolution of the uptake rate $\kappa(t)$ 
rescaled by the diffusive uptake rate $\kappa_s$
for different values of $\gamma$ (decreasing from top to bottom) 
at fixed particle radius $R=0.05$ ($\beta=D=0.01$). Resolution $M=64$.}
\label{fig:evoshear}
\end{figure}
%
Before presenting the results of simulations of
Eq.~(\ref{eq:advectiondiffusion}), we discuss the relevant scales for
well-resolving the competition between shear and diffusion.  Shear and
diffusion balance at a scale $\ell_\gamma \sim \sqrt{D/\gamma}$,
diverging for $\gamma\to 0$.  Stationarity (in the infinite volume) is
reached when the diffusive front becomes comparable with the scale
$\ell_\gamma$, i.e. for times $\tau_\gamma\sim \ell_\gamma^2/D
=\gamma^{-1}$, also diverging for $\gamma\to 0$. Thus $\ell_\gamma$
should be much smaller than the simulation box $L$ otherwise the
effect of shear starts to be effective over time scales for which the
absorber is also interacting with its periodic images. The requirement
$\ell_\gamma \ll L$ implies a constraint on the smallest shear rate
that can be used, i.e. $\gamma \gg D/L^2$. Moreover, since we are
interested in testing the prediction (\ref{eq:match}) for $Pe=\gamma
R^2/D\ll 1$, we end up with the requirements: $D/L^2 \ll \gamma \ll
D/R^2$ that can be re-expressed as $R^2/D=\tau_d \ll \tau_\gamma \ll
\tau_L=L^2/D$ and $R\ll \ell_\gamma\ll L$ in the time and scale
domain, respectively.
%
\begin{figure}[t!]
\centering
\includegraphics[width=0.49\textwidth]{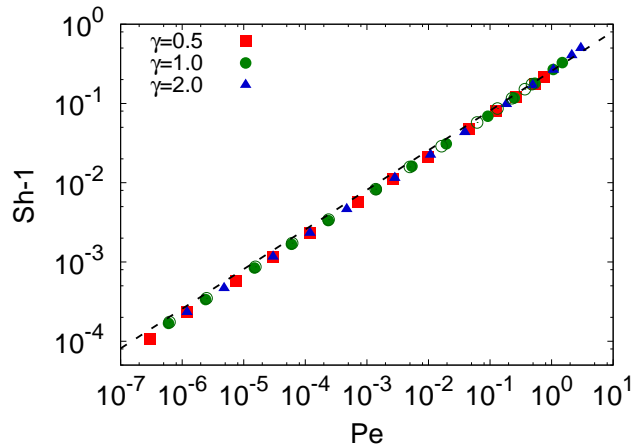}
\caption{(Color Online) Shear contribution to the Sherwood number $Sh$ 
as a function of Peclet number $Pe$ for numerical simulations at two different resolutions: 
M=64 (filled symbols) and M=128 (empty symbols). 
The line represents the theoretical behavior (\ref{eq:match}). 
The Peclet number $Pe$ is changed by fixing the shear rate $\gamma$ 
and varying the radius $R$.}
\label{fig:shear}
\end{figure}
%
The limitation on the smallest value of $\gamma$ is well evident from
Fig.~\ref{fig:evoshear}, where we show the time evolution of the
uptake rate, $\kappa(t)$, at varying $\gamma$ when $D$ and $R$ are
fixed.  For time $t \ll \tau_\gamma$ $\kappa(t)$ is essentially
indistinguishable from that obtained in the diffusive case
($\gamma=0$). For $\gamma\lesssim 0.01$, $\tau_\gamma$ is comparable
with the time at which the decay induced by the absorber periodic
images becomes effective. Figure~\ref{fig:evoshear} also show that,
due to the shear, the time behavior of the uptake rate is quite
different from the Smoluchovsky (diffusive) result,
Eq.~(\ref{eq:smolunsteady}). As a consequence, we cannot exploit
(\ref{eq:smolunsteady}) to fit the rate, as previously done. Without a
theoretical prediction for $\kappa(t)$, we can extract the (infinite
volume) uptake rate constant using Eq.~(\ref{eq:totalrate}).  Assuming
a constant uptake rate $\kappa$, the mean concentration should decay
linearly in time as $C(t) = C_0 - (\kappa/V) t$ with $V=L^3=(2\pi)^3$.
The above functional form should be fitted in the time interval
$\tau_\gamma < t < \tau_L$, when the disturbance induced by the shear
is well-developed and (quasi-)stationary.
In order to test the prediction (\ref{eq:match}) we proceed as
follows.  Given the diffusion coefficient $D$, we fixed the shear rate
$\gamma$ at three representative values, such that $\ell_\gamma$ is
well resolved by the numerical grid and $\ell_\gamma \ll L$. Then we
explored different values of Peclet number $Pe$ by varying the
particle radius $R$ (viz. the absorption rate $\beta$), but enforcing
the constraint $\ell_\gamma\gg R$ to have a small $Pe$.  We performed
two series of simulations using grid resolution $M=64,128$. For each
series of simulations we explore a sufficiently wide range of values
of Peclet number, fitting the uptake rate constant as described
above. As shown in Figure~\ref{fig:shear}, the excess Sherwood number,
$Sh-1$, as a function of $Pe$ compares very well with
(\ref{eq:match}).

\section{Conclusions}\label{sec:conclusion}

We presented a novel numerical method for computing the nutrient
uptake rate by small spherical particles immersed in a concentration
field, in the diffusion-limited regime.  The method, inspired by the
Force Coupling Method, represents each particle as an effective sink
of concentration and can in principle be used in presence of a generic
underlying flow, motile particles, and source terms for the
concentration field. Moreover, more complex reaction scheme can be
easily implemented to mimic partial or saturable absorption.\\
\indent By comparing the numerical results with exact results obtained from a
multipole expansion method based on re-expansion formulae for solid
harmonics, we have shown that the method, here implemented on a
pseudo-spectral solver, is able to correctly reproduce the diffusive
interactions among competing absorbers arranged in geometrical
configurations of increasing complexity.  As discussed the main
limitation of the method resides on the possibility to resolve
diffusive interactions at small distances, but this can be cured
increasing the resolution. Another limitation pertain to the large
distances, but this is only due to the periodicity of the simulation
domain and thus it does not concern the method itself.\\
\indent The main advantages of the method are the scalability with the number
of absorbers and the possibility to include the presence of an
arbitrary flow, for which we show a benchmark in the case of a linear
shear.  These properties make our numerical method ideal for the study
of problems possibly involving complex, turbulent flows, such as the
efficiency of nutrient uptake by microorganisms in the ocean.  In
future investigations we plan to implement the presented method to
particles transported by turbulent flows, back-reacting on it and
possibly equipped with self-propulsion.

\begin{acknowledgments}
We acknowledge the European COST Action MP1305 "Flowing Matter".
\end{acknowledgments}

\appendix
\section{Smoluchowski Formula}
\label{app1}
The problem of diffusion-limited reaction was 
first studied by Smoluchowski~\cite{smoluchowski1917} and then applied
to the heat flow into a sphere with a constant temperature
\cite{carslaw1959}. In the ecology of phytoplankton the model 
was first introduced by Osborn \cite{osborn1996}.
In the absence of a flow, the uptake by a single spherical cell 
is controlled by the diffusion equation
\begin{equation}
\partial_t \rho = D \nabla^2 \rho
\label{eq:app1.1}
\end{equation}
where D is the diffusivity and boundary conditions (for a perfect 
absorber) are $\rho=0$ at $r=R$ 
and $\rho=\rho_\infty$ as $r\rightarrow\infty$.\\
\indent Using the Laplace transform, equation (\ref{eq:app1.1}) gives the solution
\begin{equation}
\rho(r,t)=\rho_\infty \left[ 1 - \frac{R}{r} 
\erfc\left(\frac{r-R}{2\sqrt{Dt}}\right) \right]
\label{eq:app1.2}
\end{equation}
The flux per unit area is $J(r,t)=-D\partial_r \rho$. Integrated over 
the solid angle at $r=R$, this gives the rate of nutrient flux entering into 
the cell surface, i.e. $\kappa(t)=J(R,t)\,4\pi R^2$. 
Therefore, from (\ref{eq:app1.2}), the uptake rate at the sphere is
\begin{equation}
\kappa(t)=4\pi D R \rho_\infty  \left(1+\frac{R}{\sqrt{\pi D t}}\right)
\label{eq:app1.3}
\end{equation}
In the limit of long times $t\rightarrow\infty$, $\kappa(t)$ reduces to the
Smoluchowski constant rate $\kappa_s$.

\section{The mean-field theory of 
absorption by a spherical potential}
\label{app2}

\noindent Here we aim at solving the following equation:
\begin{equation}
\nabla^2 \varphi - k\Theta(b-|\bm r|) \varphi=0 \,, 
\end{equation}
with the boundary condition $\varphi(|\bm r|\to\infty)= \varphi_\infty$, 
and where $\Theta$ is the Heaviside function.
The solution we are interested in is spherically symmetric so, 
denoting  $r=|\bm r|$, the equation we actually need to solve is:
\begin{equation}
\varphi^{\prime\prime}+ 2\varphi^\prime/r - \varphi/\xi^2=0\,, 
\end{equation}
where $\xi=\sqrt{D/k}$ has dimensions of a length, and the prime denotes the derivative with respect to $r$.
The general solution is \cite{polyanin1995exact}
\begin{equation}
\varphi= \left\{
\begin{array}{ll}
\frac{C_1 \xi \sinh(r/\xi)+C_2 \xi \cosh(r/\xi)}{r}  & r\leq b\\
\frac{C_3}{r}+\varphi_\infty & r\geq b\,.
\end{array}
\right.
\label{eq:app2_solution0}
\end{equation}
To avoid a singular solution in $r=0$ we impose $C_2=0$,
while $C_1$ and $C_3$ can be fixed imposing continuity of $\varphi$ and its derivative in $r=b$.
The final result is:
\begin{equation}
\varphi (r)= \left\{
\begin{array}{ll}
\frac{\varphi_\infty \xi}{\cosh(b/\xi)} \frac{\sinh(r/\xi)}{r}  & r\leq b\\
\varphi_\infty \left(1-\frac{b-\xi\tanh(b/\xi)}{r}\right) & r\geq b\,.
\end{array}
\right.
\label{eq:app2_solution}
\end{equation}
For the results presented in the main text 
we need to compute the flux on the surface of the sphere of radius $b$, which is simply obtained as:
\begin{equation}
\kappa = 4\pi Dr^2 \left.\frac{\partial \varphi}{\partial r}\right|_{r=b}= 4\pi D\varphi_\infty\left[b-\xi\tanh(b/\xi)\right]
\label{eq:app2_rate}
\end{equation}
%

\section{The absorption by a spherical shell potential}
\label{app3}

Here we aim at solving  equation (\ref{eq:spsh}).
Similarly to the case discussed in Appendix~\ref{app2} (see
Eq.~(\ref{eq:app2_solution0}), we have three regions with different
solutions. In the interior of the shell, for $r<R_s-2R$, we have the
solution $\psi_I=a_1+b_1\xi/r$, clearly $b_1=0$ due to the divergence at
$r=0$. In the region $R_s-2R < r < R_s $ the solution is
$\psi_{II}=(\xi/r)(a_{2}e^{r/\xi}+b_{2}e^{-r/\xi})$. In
the outer region,  $r>R_s$, the solution is $\psi_{III} = a_3+b_3\xi/r$.\\
\indent The boundary condition at infinity implies that $a_3=\psi_\infty$. Imposing the
continuity of the solution and its derivative at $r=R_s-2R$ and $r=R_s$
we obtain the remaining unknown constants. The final solution is
\begin{small}
\begin{equation}
\dfrac{\psi(x)}{\psi_\infty}=
\left\{
\begin{array}{ll}
\dfrac{1}{X_1\sinh\left(X\right) \!+\! \cosh\left(X\right)} & (I) \\[.5cm]
\dfrac{1}{x}\dfrac{X_1\cosh\left(x \!-\! X_1\right)+\sinh\left(x \!-\! X_1\right)}
{X_1\sinh\left(X\right) \!+\! \cosh\left(X\right)} & (II) \\[.5cm]
1 \!-\! \dfrac{1}{x}\dfrac{\left(X_1X_2 \!-\! 1\right)\tanh\left(X\right) \!+\! X}
{X_1\tanh\left(X\right) \!+\! 1} & (III)
\end{array}
\right.
\label{eq:app3_solution}
\end{equation}
\end{small}
where  $x=r/\xi$, $X=2R/\xi$,a $X_1=(R_s-2R)/\xi$, $X_2=R_s/\xi$ 
and the three regions correspond to:  $I=[0,X_1]$, $II=[X_1,X_2]$ and $III=[X_2,\infty)$.\\
\indent As before, the total uptake rate at $r=R_s$ is given by
\begin{equation}
  \kappa_{tot} = 4 \pi D r^2\left.\frac{\partial \psi}{\partial r}\right|_{r=R_s}=4\pi D b_3 \xi \,,
  \label{eq:app3_rate}
\end{equation}
where $b_3$ can be read from the term proportional to $1/x$ in
(\ref{eq:app3_solution}).III.

\bibliography{biblio_short}

\end{document}